\documentclass[
amsmath, %
floatfix, %
twocolumn, %
reprint, %
superscriptaddress, %
prl, %
aps, %
citeautoscript, %
final, %
]{revtex4-2} 
\renewcommand{\thispagestyle}[1]{}

\usepackage[T1]{fontenc}
\usepackage[utf8]{inputenc}
\usepackage{graphicx}
\usepackage{latexsym}
\usepackage{amsthm}
\usepackage[low-sup]{subdepth}

\usepackage{color}
\usepackage{mathtools}
\mathtoolsset{showonlyrefs}

\usepackage{placeins}

\usepackage[dvipsnames,table]{xcolor}

\usepackage[]{newtxtext} 
\usepackage[subscriptcorrection,nosymbolsc,smallerops,bigdelims]{newtxmath} 
\DeclareMathAlphabet{\mathcal}{OMS}{cmsy}{m}{n} 
\DeclareMathAlphabet{\mathbcal}{OMS}{cmsy}{b}{n} 
\usepackage{bm}

\usepackage{floatrow}
\usepackage{siunitx}
\usepackage{hyperref}
\hypersetup{
	colorlinks,
	linkcolor={blue!90!black},
	citecolor={blue!90!black},
	urlcolor=	{blue!90!black}
}

\usepackage{multirow}

\let\oldeqref\eqref
\renewcommand*{\eqref}[1]{%
	\hyperref[#1]{\oldeqref{#1}}%
}

\renewcommand\Im{\operatorname{Im}}

\mathchardef\mhyphen="2D

\DeclarePairedDelimiter\lr{\lparen}{\rparen}

\DeclarePairedDelimiter\abs{\lvert}{\rvert}

\DeclarePairedDelimiterX{\comm}[2]{\lbrack}{\rbrack}{#1, #2}
\DeclarePairedDelimiter\ket{\lvert}{\rangle}
\DeclarePairedDelimiter\bra{\langle}{\rvert}
\DeclarePairedDelimiterX{\braket}[2]{\langle}{\rangle}{#1\delimsize\vert #2}
\DeclarePairedDelimiterX{\ketbra}[2]{\rvert}{\lvert}{#1 \delimsize\rangle\!\delimsize\langle #2}
\DeclarePairedDelimiterX{\matrixel}[3]{\langle}{\rangle}{#1 \delimsize\vert #2 \delimsize\vert #3}

\makeatletter
\newcommand{\raisemath}[1]{\mathpalette{\raisem@th{#1}}}
\newcommand{\raisem@th}[3]{\raisebox{#1}{$#2#3$}}
\makeatother

\definecolor{cbred}{HTML}{e31a1c}
\definecolor{cbgreen}{HTML}{33a02c}
\definecolor{cbblue}{HTML}{176aa7}

\usepackage{dcolumn}
\newcolumntype{d}[1]{D{.}{.}{#1}}

\newcommand{\figref}[1]{Fig.~\ref{fig:#1}}
\newcommand{\afigref}[1]{Figure~\ref{fig:#1}}
\newcommand{\subfigref}[2]{Fig.~\hyperref[fig:#1]{\ref*{fig:#1}(#2)}}
\newcommand{\asubfigref}[2]{Figure~\hyperref[fig:#1]{\ref*{fig:#1}(#2)}}
\newcommand{\subfigsref}[3]{Figs.~\hyperref[fig:#1]{\ref*{fig:#1}(#2)}-\hyperref[fig:#1]{\ref*{fig:#1}(#3)}}

\definecolor{cbred}{HTML}{e31a1c}
\definecolor{cbgreen}{HTML}{33a02c}
\definecolor{cbblue}{HTML}{176aa7}
\definecolor{cborange}{HTML}{ff7f00}
\definecolor{cbviolet}{HTML}{6a3d9a}

\usepackage[activate={true,nocompatibility},final,tracking=alltext,kerning=true,spacing=true,protrusion=true,factor=1100,stretch=10,shrink=10,selected=true ,letterspace=-0]{microtype}

\usepackage{array,ragged2e}
\usepackage{orcidlink}

\begin{document}
	\title{Optical Activity of Group III-V Quantum Dots Directly Embedded in Silicon}
	
	\author{M.~Gawe{\l}czyk}
	\affiliation{Institute of Theoretical Physics, Wroc\l{}aw University of Science and Technology, 50-370 Wroc\l{}aw, Poland}

	\author{K.~Gawarecki}
	\affiliation{Institute of Theoretical Physics, Wroc\l{}aw University of Science and Technology, 50-370 Wroc\l{}aw, Poland}

	\begin{abstract}
		Optically active III-V group semiconductor quantum dots (QDs) are the leading element of the upcoming safe quantum communication. However, the entire electronic and IT infrastructure relies on silicon-based devices, with silicon also providing a natural platform for photonic integration. Combining semiconductor optics with silicon electronics is thus a major technological challenge. This obstacle cannot be directly solved because silicon is optically inactive. Interfacing III-V quantum dots with silicon is thus a sought-after solution.
		A radical approach is to embed III-V material grains directly into silicon. The first realization of such technology was developed, and it gave InAs and core-shell InAs/GaAs QDs embedded in Si with bright and narrow single-QD emission lines. No theory has been given, though, and, as we show here, it is not even obvious if and how such QDs can be optically active.
		We first use general arguments, also supported by atomistic calculations, that InAs/Si QDs cannot confine both carrier types unless the structural strain is mostly relaxed, meaning many defects at the interface. This explains the lack of light emission from those dots. Then we show that the InAs/GaAs/Si QDs can confine both carrier types. Their electron states are, however, highly influenced by $k$-space valley mixing, which impacts emission spectra and deteriorates optical properties. We propose to overcome this by adding an additional wider-bandgap material layer.
	\end{abstract}
	
	\maketitle

	Single-photon emitters are fundamental for quantum information technologies such as secure communication \cite{Aharonovich2016,Wehner2018}. However, for such technologies to be actually useful, scalable quantum photonic systems need to be realized. For those, it is essential to integrate both the photon emitters and other passive components of the system into on-chip quantum photonic integrated circuits. Silicon is a well-established platform for such photonic integration thanks to its low optical losses at telecom wavelengths and compatibility with the existing technology. However, as an indirect bandgap semiconductor, it fundamentally does not emit light, so other sources need to be integrated into silicon devices.

    Among many possible approaches, the most explicit one is to directly integrate III–V semiconductor quantum dots (QDs) in Si matrices \cite{Reithmaier2018}. If successful, this method would offer a pathway toward compact and technology-compatible single-photon sources. In particular, III–V QDs like InAs-based systems exhibit excellent optical characteristics, including emission at telecom wavelengths. Nevertheless, heteroepitaxial growth of III–V materials on silicon presents substantial challenges, primarily due to enormous lattice mismatch, thermal expansion coefficient differences, and interfacial defects\cite{Kunert2018}. All these factors can contribute to the degradation of emitter performance.

    Despite these obstacles, recent progress in epitaxial growth techniques has brought such monolithic integration of InAs grains in Si \cite{Benyoucef2013,Benyoucef2013b,Wu2015,Reithmaier2018}. However, obtaining emission turned out to be not straightforward, and required using a core-shell structure of the embedded grains. Here, we focus on those III-V QDs directly embedded in Si and first answer the question of why InAs/Si structures have never been shown to emit light. Such QDs either have a defect-rich interface, obviously obstructing optical activity, or are highly strained. For the latter case, we find that those objects do not confine both carrier types due to band alignment, which is strongly impacted by deformation potentials. Then, we switch to InAs/GaAs/Si QDs that have been shown to emit \cite{Benyoucef2013} and, based on our considerations, are also more promising for providing enough confining potential for electrons and holes. However, electrons tend to be confined in the GaAs shell, and their confined states heavily hybridize with a dense ladder of (partly) extended X-valley states. This mixing spoils QD optical properties by smearing the oscillator strength over a quasi-continuum of such hybridized exciton states. The resulting spectra are rich and interesting, but do not promise excellent performance as non-classical light sources in quantum technologies. Thus, we propose to cut off the hybridization by adding one more layer of semiconductor that provides a wider bandgap not only at the $\Gamma$ point but importantly in the X-valley.
	
    \begin{figure}[tb] %
	\begin{center} %
		\includegraphics[width=\columnwidth]{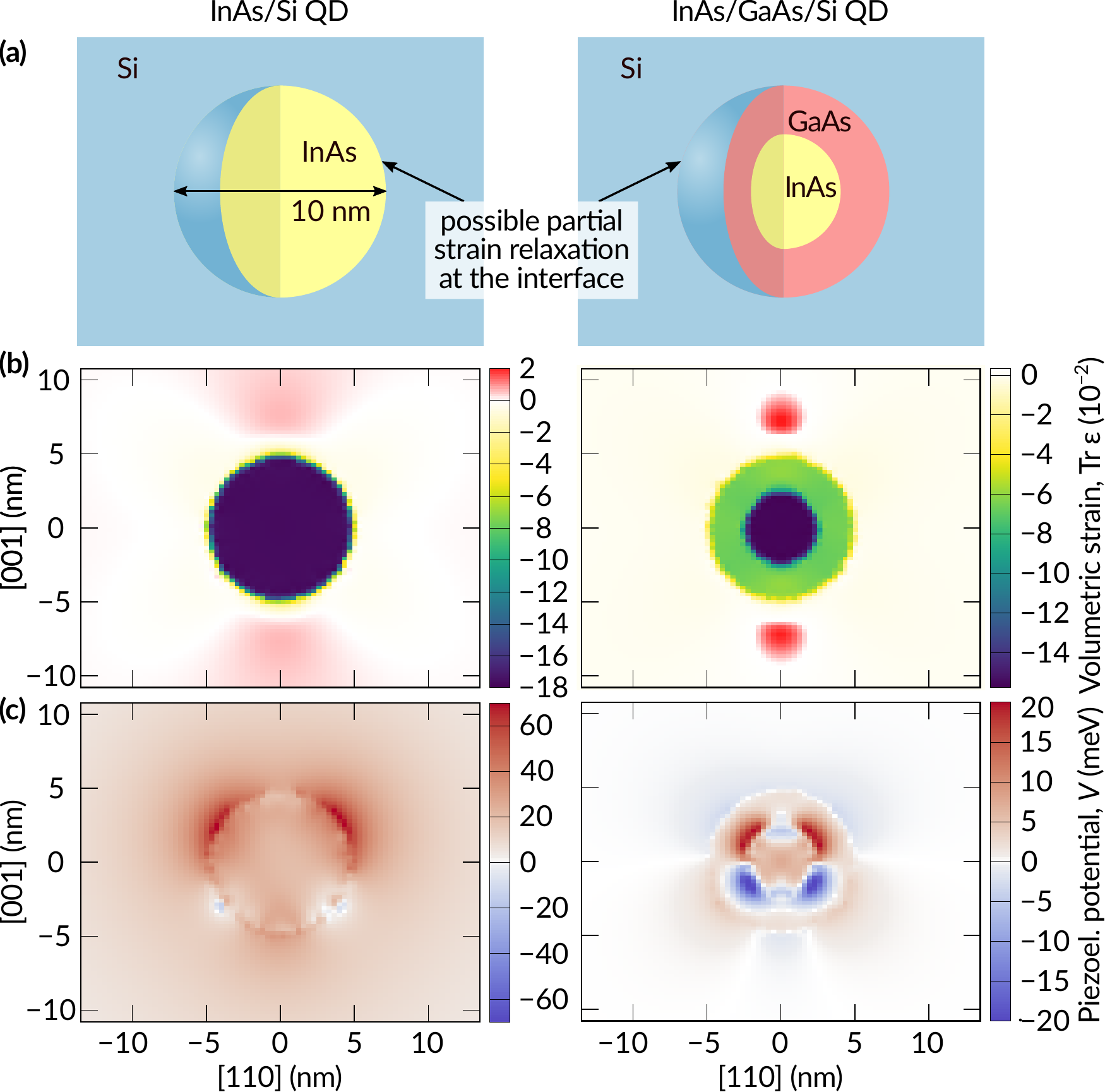} %
	\end{center} %
	\caption{\label{fig:struct-strain}{\bf Simulated III-V semiconductor quantum dots directly embedded in the silicon matrix.} \textbf{(a)} Schematic presentation of the InAs/Si (left) and InAs/GaAs/Si (right) QDs. \textbf{(b)} Volumetric strain in the structures assuming ideal, defect-free interfaces. \textbf{(c)} Piezoelectric potential caused by the shear strain in piezoelectric III-V materials.} %
	\end{figure}	

	In \subfigref{struct-strain}{a}, we schematically show the two types of QDs that resemble those that have been realized technologically \cite{Benyoucef2013,Benyoucef2013b,Wu2015,Reithmaier2018} and are studied here: InAs clusters directly embedded in Si (left), and core-shell InAs/GaAs QDs in Si (right). For simplicity, we assume full spherical symmetry of the dots. In each of the cases, the interface with Si can feature lattice defects \cite{Benyoucef2013}, for which we further account by optionally relaxing the strain, completely or partially. Given that such interfacial defects are detrimental to optical properties, we mainly focus on the highly strained defect-free case.

    \begin{figure*} %
	\begin{center} %
		\includegraphics[width=\textwidth]{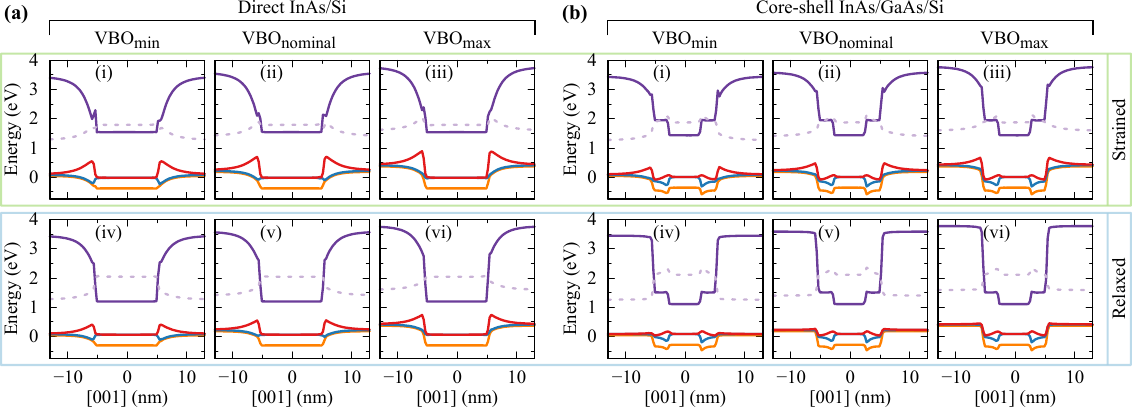} %
	\end{center} %
	\caption{\label{fig:band-edges}{\bf Calculated band edges along the [001] axis for III-V semiconductor quantum dots in the silicon matrix.} Solid lines mark the bands at the $\Gamma$ point of the Brillouin zone, dashed line shows the conduction band edge at $\bm{k}_0$ points corresponding to conduction band minima in Si (X-valleys). \textbf{(a)} InAs/Si QDs; Panels (i)-(iii) (top row) show results for the fully strained system, while (iv)--(vi) (bottom row) assume full relaxation of strain at the interface with Si. Columns correspond to different valence band offset (VBO) values of Si with respect to GaAs, based on the literature. \textbf{(b)} As in (a) but for the core-shell InAs/GaAs/Si QDs.} %
    \end{figure*}
	
	To account for the significant lattice mismatch and resulting strain, we use the atomistic valence force field model of Keating \cite{Keating1966} and find the atom positions minimizing the strain energy of the system (see Appendix for details). For presentation these atom displacements are translated to the continuum model \cite{pryor98b} with the strain tensor elements $\varepsilon_{ij}=(\partial u_i / \partial x_j +\partial u_j / \partial x_i)$, where $\bm{u}$ is the vector displacement field and $x_i$ are the Cartesian coordinates coinciding with the [100], [010] and [001] crystallographic axes. \asubfigref{struct-strain}{b} shows the most important strain contribution in a defect-free QD, which is the volumetric strain $\mathrm{Tr\widehat{\varepsilon}}$ that measures the relative change of the unit cell volume in the system. Via deformation potential, this contribution will heavily shift the band edges and thus change the bandgap energy in a QD. Spherical symmetry of the modeled QDs leads to no biaxial strain. However, inclined or curved interfaces always lead to the appearance of shear strain, which in turn causes piezoelectric polarization in non-centrosymmetric materials. Here, only the III-V materials exhibit this effect, which is enough to produce a significant piezoelectric field inside a QD shown in \subfigref{struct-strain}{c}.

	To find the states of electrons and holes in the QDs, we use the calculated atomic positions in the empirical tight-binding (TB) method with the nearest-neighbor approximation, including sp$^3$d$^5$s$^*$ orbitals~\cite{Slater1954,Jancu1998,zielinski12}, spin-orbit interaction~\cite{Chadi1977}, and strain \cite{Ren1982,Jancu1998,Jancu2007,Gawarecki2019b} (see Appendix for details and material parameters used). On top of that, we calculate the excitonic states, and compute their optical properties in the dipole approximation~\cite{Feynman1939, LewYanVoon1993, eissfelleer12, Gawarecki2020}, which yields their oscillator strength and related radiative lifetimes, and allows us to simulate the absorption spectra (see Appendix for details). As it will be shown, due to the energetic vicinity of the QD confined electron states and bulk Si conduction-band minima in the X-valleys (at $\bm{k}_0 \simeq 0.85 (2\pi/a) [1,0,0]$ and equivalent points close to the X point), we need to work here with very high numbers (up to a few hundred) of single-particle electron states. This impedes the usage of the full configuration-interaction method \cite{Bryant1987} to account for electron-hole interactions. For that reason, we limit the calculation to only include diagonal couplings \cite{Gawarecki2024}, which account for most of the electron-hole interaction energy, but neglect the effects of charge redistribution and correlation. While significant methodological development is needed to describe excitonic effects in such a system more accurately, our approach is enough to grasp the main optical properties of the system in this initial study.

	Before we focus on the eigenstates, the band edges themselves in the system deserve attention and provide an explanation for a considerable part of its features. In \figref{band-edges}, we present the calculated band edges, resulting from our TB simulation. \asubfigref{band-edges}{a} shows the results for the InAs/Si QD, while \subfigref{band-edges}{b} for the InAs/GaAs/Si one. In each case, the top row presents the band edges in the defect-free system that is highly strained, while the bottom one shows the opposite extreme case with the strain fully relaxed, presumably due to interface defects. While most of the parameters for the constituent materials are well-established (see Appendix), the valence-bend offset (VBO) between Si and III-V materials has a large spread in the literature \cite{Choi2013,Krawicz2014,Kang2015}. For that reason, we show in each case three columns of panels, with the middle one resulting from the VBO value we treat as nominal (VBO$_{\mathrm{nominal}}=167.5$~meV above GaAs \cite{Choi2013}) and the left/right ones corresponding to the extreme cases of VBO$_{\mathrm{min}}=90$~meV and VBO$_{\mathrm{max}}=420$~meV, respectively. 

	In each of the panels, solid lines mark the band edges at the $\Gamma$ point, while the dashed lines additionally show the conduction band edge at the $\bm{k}_0$ points where Si has the conduction band minima. Red lines mark the heavy-hole band, and one can notice that the system generally provides a type-II confinement with a confining potential for the hole at the interface. In the conduction band, we obtain a well-defined potential well at the $\Gamma$ point. Despite this, obtaining electron confinement is not straightforward in the system, as the $\Gamma$ point band-edge minima lie close to or even above the bulk level of the Si conduction band at the $\bm{k}_0$ points. Comparing the strained and unstrained cases, we can infer that this issue can be partly resolved when strain is reduced. However, this comes at the expense of making the valence-band confinement even shallower. The core-shell system [\subfigref{band-edges}{b}] seems more promising for the electron confinement, as the GaAs shell creates a potential barrier also in the X-valleys. However, one has to take the difference in effective masses between the materials into account while studying the presented diagrams. Slightly deeper confining potential in the InAs core, where electrons are lighter, $m^*=0.023 m_0$, may not be enough to confine the electron in view of its much higher effective mass in GaAs ($0.063 m_0$). In addition, this has to be compared with an order of magnitude higher effective mass in Si ($0.26 m_0$ average over anisotropy). Thus, already based on \subfigref{band-edges}{b} and having the effective masses in mind, we may conclude that the InAs/Si system either does not provide electron confinement in the strained case, or needs almost full strain relaxation to provide it. This, in turn, means numerous defects present directly at the interface where hole confinement is present. This explains the lack of emission from this kind of QDs in the experiment \cite{Benyoucef2013b}.
	
	A similar reasoning has lead to the fabrication of the core-shell InAs/GaAs/Si QDs \cite{Benyoucef2013b}, with the argumentation that the shell should mainly act as a layer keeping the interface defects further from the InAs core serving as the confining region for the carriers. Inspecting \subfigref{band-edges}{b}, we see that again we deal here with a compromise between electron and hole confinement in the relaxed or strained systems. For further study, we select the case of a fully strained QD as the limiting case, promising the best optical properties. We note, however, that the optimal outcome would be to not only obtain optical activity but also tune it to telecom wavelengths close to the 1.5~\textmu{}m low-loss transmission window of optical fibers. We leave this aspect for further studies. We also use the nominal value of the VBO from now on.
	
    \begin{figure}[tb] %
	\begin{center} %
		\includegraphics[width=\columnwidth]{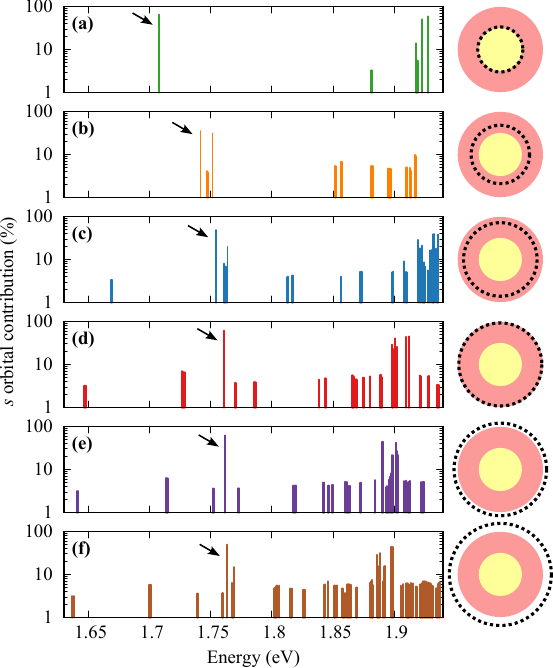} %
	\end{center} %
	\caption{\label{fig:electron-states}{\bf Calculated spectra of electron eigenstates with their \textit{s}-orbital contribution for the InAs/GaAs/Si QD.} Panels \textbf{(a)}--\textbf{(f)} show the spectra for consecutively increasing size of the computational spherical box for TB diagonalization, as marked graphically to the right of each panel. Arrows mark the most typical bound state composed mainly of the $s$ orbitals.} %
    \end{figure}
	
	As the vicinity of the indirect Si conduction band minimum seems to be the most problematic issue for the carrier confinement in the system, we now focus mainly on electron states. \afigref{electron-states} shows the spectra of electron eigenstates calculated in the strained InAs/GaAs/Si QD. The quasi-continuum of bulk Si states constitutes a significant computational challenge. For that reason, we are unable to diagonalize the TB Hamiltonian in a computational box large enough to consider the computation converged. In view of that, we adopt a different approach, in which we gradually increase the box size, to see how the Si states emerge and affect the QD-confined states. Note that the atomic positions are always found using a large computational box, and this methodology only applies to diagonalizing the TB Hamiltonian in a reduced box. Such results are shown in consecutive rows of \figref{electron-states}, where we begin with a box encompassing only the InAs core of the QD, and finish with the one containing some of the Si bulk material. The arrows mark the lowest confined states as they evolve with the increasing box size. The bar heights show the $s$-orbital contribution to the states, with the conduction states at the $\Gamma$ point expected to be predominantly composed of these orbitals. One can notice that with the increasing box size, a large number of other states appear, which are related to the vicinity of $\bm{k}_0$ points either in the GaAs shell or further in the Si. Each time the QD confined states fall close in energy with one of those states, they hybridize, which can be seen from the decreased $s$-orbital contribution to the state. At this point, we need to note that in a large box we would deal with a quasi-continuum of those X-valley states, which would hybridize with the confined states of interest. While this should not completely destroy the optical transitions, it will smear them over a range of states into which the confined states will evolve due to hybridization.
	
    \begin{figure}[tb] %
	\begin{center} %
		\includegraphics[width=\columnwidth]{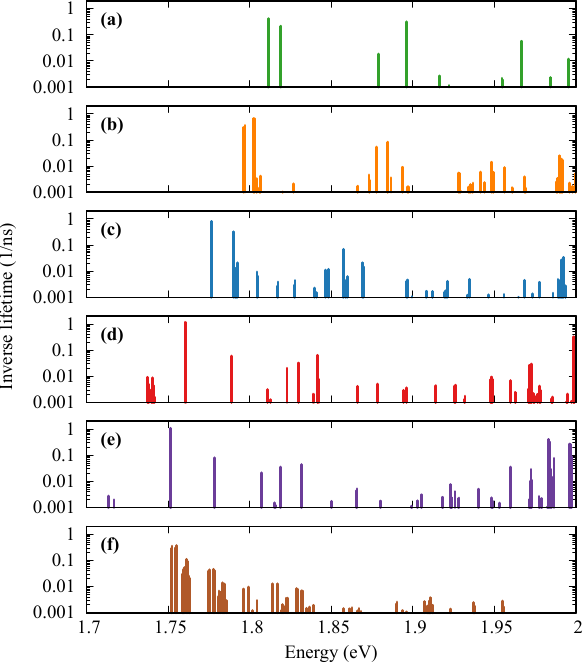} %
	\end{center} %
	\caption{\label{fig:oscillator-strength}{\bf Spectra of neutral exciton state inverse lifetimes for the strained InAs/GaAs/Si QD.} As in \figref{electron-states}, panels \textbf{(a)}--\textbf{(f)} show the spectra for consecutively increasing size of the computational spherical box used for TB diagonalization.}
    \end{figure}
	
	Having noticed that, we move to the optical properties of strained InAs/GaAs/Si QDs. \afigref{oscillator-strength} shows the spectra of exciton states, with the bar height indicating the inverse of the state radiative lifetime. Again, consecutive rows of plots correspond to increasing computational box size in TB diagonalization and show how the initially sparse spectrum of bright states confined in the QD core evolves upon the addition of the GaAs shell states and states connected to the Si X-valley. We note how the lowest transitions that occur at around 1.82~eV are subject to a redshift, but at the same time, they evolve into a large number of closely lying states with decreased optical activity, evidenced by at least a few times longer radiative lifetimes. Again, this hybridization would further evolve into a quasi-continuum in a sufficiently large computational box. Thus, while the system is bright in principle, it does not provide well-defined spectrally isolated transitions. This stands in the way of applying such QDs as quantum emitters, e.g., in scenarios requiring single-photon or entangled-photon pair generation.

	Thus, strong hybridization with the X-valley electron states is not merely a computational challenge but also forms an obvious problem for the use of III-V/Si QDs as nonclassical light sources. Therefore, to circumvent this issue, it is necessary to cut off the coupling to those extended states. A promising pathway could be to either replace the GaAs shell, or to add another layer of material that provides a wide bandgap both at the $\Gamma$ but also in the X-valley. While the QD states would still energetically overlap with those of Si, such a potential barrier would limit their hybridization and thus lead to cleaner optical spectra.

	In conclusion, we have studied the III-V semiconductor QDs directly embedded in silicon and focused on their optical properties. Our results show that obtaining a well-defined confinement for both electrons and holes in such structures is not straightforward. In the strained InAs/Si system, the direct vicinity of the X-valley silicon conduction states prevents trapping electrons in a QD, which is another explanation for the lack of emission from such structures, in addition to defects at the interface in the case of relaxed structure. In the core-shell InAs/GaAs/Si structure, local confining potential minima can be obtained for both carrier types. Yet, the electrons tend to be confined in the GaAs shell, and again are subject to strong hybridization with the X-valley silicon states. This not only spoils the optical properties but also makes a fully converged numerical simulation infeasible. Nonetheless, we have shown a series of simulations with an increasing computational box size, which demonstrate how the coupling to silicon conduction-band minimum states impacts the simulated optical spectra, smearing the oscillator strength over numerous closely spaced transitions.
	
	While this study provides an initial, limited insight into the optical activity of III-V/Si quantum dots, it serves as a crucial first step towards a comprehensive understanding and eventual engineering of their excitonic and optical properties. On the theory side, methods that would explicitly handle the silicon continuum states alongside discrete III-V QD confined states and include their mutual coupling need to be developed to address the limitations of the present work. Regarding technology and applications, we anticipate a variety of growth modifications that could limit the coupling to silicon conduction states that may eventually provide good-quality quantum emitters.

	\begin{acknowledgments}
		We acknowledge support from the National Science Centre (Poland) under Grant No. 2019/33/B/ST5/02941.
		Numerical calculations have been carried out using resources provided by Wroclaw Centre for Networking and Supercomputing \cite{wcss}, Grant No.~203.
		We are grateful to Grzegorz S\k{e}k and Mikhail Nestoklon for helpful discussions.
	\end{acknowledgments}
	\FloatBarrier

\appendix
\section*{Appendix: Calculation details}

\subsection{Calculation of strain}
\label{app:minstrain}

\begin{table}
	\caption{\label{tab:params_strain} The material parameters: the lattice constant $a$ (in \AA), the bulk elastic constants (in $10^3$~dyne/cm$^2$)~\cite{pryor98b,vurgaftman01}, and the force constants $\alpha$ and $\beta$ (in $10^{12}$~dyne/cm).}
	\begin{ruledtabular}
		\begin{tabular}{lllccc}
			& InAs & GaAs & AlAs & Si\\
			\hline \\[-0.5em]
			$a$ & 6.0583 & 5.65325 & 6.071 & 5.429 \\[1.1pt]
			$C_{11}$  & 8.53 & 12.21 & 9.63 & 16.57 \\[1.1pt]
			$C_{12}$  & 4.90 & 5.66 & 4.62 & 6.39 \\[1.1pt]
			\hline\\[-0.5em]
			$\alpha$  & 35.18 & 41.25 & 40.36 & 48.51 \\[1.1pt]
			$\beta$   &  5.49 &  9.26 & 10.13 & 13.82 \\[1.1pt]
			
		\end{tabular}
	\end{ruledtabular}
\end{table}

The lattice mismatch between the materials constituting the nanostructure inevitably leads to strain in the system. To account for this effect, we utilize the Keating Valence Force Field (VFF) model, where we minimize the elastic energy in the form~\cite{pryor98b}
\begin{align*}
	U_{\mathrm{K}} &= \frac{3}{16} \sum_{i} \sum_{j}^{\mathrm{NN}(i)} \frac{ \alpha_{ij} }{d_{ij}^{2}} \left( \lvert\bm{R}_j - \bm{R}_i\rvert^2 - d_{ij}^{2} \right)^{2} \\
	& \phantom{=} + \frac{3}{16} \sum_{i}  \sum_{j,k>j}^{\mathrm{NN}(i)}  \frac{ \tilde{\beta}_{ijk} }{d_{ij} d_{ik}}  \Big [ (\bm{R}_j - \bm{R}_i) \cdot (\bm{R}_k - \bm{R}_i) \\ &  \phantom{=} - d_{ij} d_{ik} \cos{\theta_{0}} \Big ]^{2},	
\end{align*}
where $d_{ij}$ are the ideal (unstrained) bond lengths between the atoms (the node $i$ and $j$), $\bm{R}_i$ are the actual atomic positions, and $\cos{\theta_0}$ is the ideal bond angle (which obeys $\cos{\theta_0} = -1/3$). The $\alpha$ and $\beta$ are the parameters that scale the contributions coming from changes in the bond lengths and the bond angles, respectively. The latter is averaged over the two involved atomic pairs
\begin{equation*}
	 \tilde{\beta}_{ijk} = \lr{\beta_{ij} + \beta_{ik}}/2.
\end{equation*}
For the zinc-blende lattice (which is the case here), the unstrained interatomic distance is given by $d_{ij} = a/\sqrt{3}$, where $a$ is the lattice constant corresponding to the bulk defined by a given pair of atoms. 

In the case of InAs, GaAs, AlAs, and Si, we took the values of $\alpha$ and $\beta$ reproducing the elastic constants $C_{11}$ and $C_{12}$ via $\alpha = a (C_{11} + 3 C_{12} )/4$ and $\beta = a(C_{11}-C_{12})/4$ ~\cite{Keating1966}. The value of $C_{44}$ is not fitted, as the shear strain is less important for the considered system~\cite{Zielinski2012}. The values of parameters are given in Table~\ref{tab:params_strain}. In the case of XSi compounds, due to the lack of material parameters, we calculate $\alpha$ and $\beta$ taking the average from the values of the constituent materials.

We numerically found the atomic positions, which minimize the $U_{\mathrm{K}}$ functional. This has been done using the PETSC TAO library~\cite{petsc-web-page}.

\subsection{Tight binding model}
\label{app:tb}

We calculated the electron and hole single-particle states using the sp$^3$d$^5$s$^*$ tight-binding model~\cite{Slater1954,Jancu1998,zielinski12}. We use the nearest neighbors (NN) approximation. 
The spin-orbit coupling is taken into account via the $p$-shell on-site matrix elements~\cite{Chadi1977}.  
To avoid the surface states in the band-gap, we utilize the passivation scheme following Ref.~\cite{Lee2004}.
We took the InAs, GaAs, AlAs, and Si material parameters from Ref.~\cite{Jancu1998}. As for the strain calculations, we take the average in the case of the values for XSi compounds.

The strain modifies the bond lengths and the bond angles between the atoms in the system. The first effect is accounted for in the tight-binding Hamiltonian by rescaling the hopping (two-center) parameters $t^{(ij)}_{\alpha \beta} \rightarrow t^{(ij)}_{\alpha \beta} (d_{ij}/\abs{\bm{r}_{ij}})^{n^{(ij)}_{\alpha\beta\kappa}}$, where indices $\alpha$ and $\beta$ denote the atomic orbitals, and $n^{(ij)}_{\alpha\beta\kappa}$ are called the strain exponents~\cite{Ren1982,Jancu1998,Jancu2007,Gawarecki2019b}. The effect of the modified bond angles is inherently present in the model, as the tight-binding Hamiltonian contains the atomic positions. Following Ref.~\cite{Jancu1998}, we also incorporate a strain-induced shift to the $d$-shell on-site energy (scaled by the $b_{\mathrm{d}}$ parameter).

\subsection{The optical activity}
\label{sec:optical}

The absorbance spectrum can be calculated from the imaginary part of the dielectric function 
\begin{equation}
	\label{eq:absPropTo}
	\alpha(\omega) \propto {\omega \Im{\epsilon(\omega)}} \propto \sum_{n} \frac{f_{n} \, \omega \, \gamma}{(\omega^2_{n} - \omega^2)^2 + \omega^2 \gamma^2},
\end{equation}	 
where $\gamma$ is the decrement parameter (representing the line broadening), $f_{n}$ are the exciton oscillator strengths given by
\begin{equation}
	f_{n} = \frac{2}{m_0 E^{(\mathrm{X})}_n} \sum_{i=x,y,z} \abs{\bra{\mathrm{vac.}}{P_i}\ket{X_n}}^2 ,
\end{equation}
where $E^{(\mathrm{X})}_n = \hbar \omega_n$ is the energy of the exciton state $\ket{X_n}$. Due to the large number of states, the full configuration-interaction calculations are not feasible. In consequence, we include the Coulomb interaction only via the diagonal terms.
The momentum matrix elements are calculated using the Hellmann-Feynman theorem\cite{Feynman1939, LewYanVoon1993, eissfeller12,Gawarecki2020}. 

The radiative decay rate of the $\ket{X_n}$ state is calculated using~\cite{Thranhardt2002,Gawelczyk2017}
\begin{align*}
    \tau^{-1}_n &= \frac{n_r(E^{\mathrm{(X)}}_n/\hbar) \lr{E^{\mathrm{(X)}}_n}^2 f_n}{6 \pi \epsilon_0 m_0 c^3  \hbar^2},
\end{align*}
where $n_r(\omega)$ is the refractive index.


\begin{thebibliography}{31}%
	\makeatletter
	\providecommand \@ifxundefined [1]{%
		\@ifx{#1\undefined}
	}%
	\providecommand \@ifnum [1]{%
		\ifnum #1\expandafter \@firstoftwo
		\else \expandafter \@secondoftwo
		\fi
	}%
	\providecommand \@ifx [1]{%
		\ifx #1\expandafter \@firstoftwo
		\else \expandafter \@secondoftwo
		\fi
	}%
	\providecommand \natexlab [1]{#1}%
	\providecommand \enquote  [1]{``#1''}%
	\providecommand \bibnamefont  [1]{#1}%
	\providecommand \bibfnamefont [1]{#1}%
	\providecommand \citenamefont [1]{#1}%
	\providecommand \href@noop [0]{\@secondoftwo}%
	\providecommand \href [0]{\begingroup \@sanitize@url \@href}%
	\providecommand \@href[1]{\@@startlink{#1}\@@href}%
	\providecommand \@@href[1]{\endgroup#1\@@endlink}%
	\providecommand \@sanitize@url [0]{\catcode `\\12\catcode `\$12\catcode
		`\&12\catcode `\#12\catcode `\^12\catcode `\_12\catcode `\%12\relax}%
	\providecommand \@@startlink[1]{}%
	\providecommand \@@endlink[0]{}%
	\providecommand \url  [0]{\begingroup\@sanitize@url \@url }%
	\providecommand \@url [1]{\endgroup\@href {#1}{\urlprefix }}%
	\providecommand \urlprefix  [0]{URL }%
	\providecommand \Eprint [0]{\href }%
	\providecommand \doibase [0]{https://doi.org/}%
	\providecommand \selectlanguage [0]{\@gobble}%
	\providecommand \bibinfo  [0]{\@secondoftwo}%
	\providecommand \bibfield  [0]{\@secondoftwo}%
	\providecommand \translation [1]{[#1]}%
	\providecommand \BibitemOpen [0]{}%
	\providecommand \bibitemStop [0]{}%
	\providecommand \bibitemNoStop [0]{.\EOS\space}%
	\providecommand \EOS [0]{\spacefactor3000\relax}%
	\providecommand \BibitemShut  [1]{\csname bibitem#1\endcsname}%
	\let\auto@bib@innerbib\@empty
	\bibitem [{\citenamefont {Aharonovich}\ \emph {et~al.}(2016)\citenamefont
		{Aharonovich}, \citenamefont {Englund},\ and\ \citenamefont
		{Toth}}]{Aharonovich2016}%
	\BibitemOpen
	\bibfield  {author} {\bibinfo {author} {\bibfnamefont {I.}~\bibnamefont
			{Aharonovich}}, \bibinfo {author} {\bibfnamefont {D.}~\bibnamefont
			{Englund}},\ and\ \bibinfo {author} {\bibfnamefont {M.}~\bibnamefont
			{Toth}},\ }\bibfield  {title} {\bibinfo {title} {{Solid-state single-photon
				emitters}},\ }\href {https://doi.org/10.1038/nphoton.2016.186} {\bibfield
		{journal} {\bibinfo  {journal} {Nat. Photonics}\ }\textbf {\bibinfo {volume}
			{10}},\ \bibinfo {pages} {631} (\bibinfo {year} {2016})}\BibitemShut
	{NoStop}%
	\bibitem [{\citenamefont {Wehner}\ \emph {et~al.}(2018)\citenamefont {Wehner},
		\citenamefont {Elkouss},\ and\ \citenamefont {Hanson}}]{Wehner2018}%
	\BibitemOpen
	\bibfield  {author} {\bibinfo {author} {\bibfnamefont {S.}~\bibnamefont
			{Wehner}}, \bibinfo {author} {\bibfnamefont {D.}~\bibnamefont {Elkouss}},\
		and\ \bibinfo {author} {\bibfnamefont {R.}~\bibnamefont {Hanson}},\
	}\bibfield  {title} {\bibinfo {title} {Quantum internet: A vision for the
			road ahead},\ }\href {https://doi.org/10.1126/science.aam9288} {\bibfield
		{journal} {\bibinfo  {journal} {Science}\ }\textbf {\bibinfo {volume}
			{362}},\ \bibinfo {pages} {eaam9288} (\bibinfo {year} {2018})}\BibitemShut
	{NoStop}%
	\bibitem [{\citenamefont {Reithmaier}\ and\ \citenamefont
		{Benyoucef}(2018)}]{Reithmaier2018}%
	\BibitemOpen
	\bibfield  {author} {\bibinfo {author} {\bibfnamefont {J.~P.}\ \bibnamefont
			{Reithmaier}}\ and\ \bibinfo {author} {\bibfnamefont {M.}~\bibnamefont
			{Benyoucef}},\ }\bibinfo {title} {Iii–v on silicon nanocomposites},\ in\
	\href {https://doi.org/10.1016/bs.semsem.2018.08.004} {\emph {\bibinfo
			{booktitle} {Silicon Photonics}}}\ (\bibinfo  {publisher} {Elsevier},\
	\bibinfo {year} {2018})\ p.\ \bibinfo {pages} {27–42}\BibitemShut {NoStop}%
	\bibitem [{\citenamefont {Kunert}\ \emph {et~al.}(2018)\citenamefont {Kunert},
		\citenamefont {Mols}, \citenamefont {Baryshniskova}, \citenamefont {Waldron},
		\citenamefont {Schulze},\ and\ \citenamefont {Langer}}]{Kunert2018}%
	\BibitemOpen
	\bibfield  {author} {\bibinfo {author} {\bibfnamefont {B.}~\bibnamefont
			{Kunert}}, \bibinfo {author} {\bibfnamefont {Y.}~\bibnamefont {Mols}},
		\bibinfo {author} {\bibfnamefont {M.}~\bibnamefont {Baryshniskova}}, \bibinfo
		{author} {\bibfnamefont {N.}~\bibnamefont {Waldron}}, \bibinfo {author}
		{\bibfnamefont {A.}~\bibnamefont {Schulze}},\ and\ \bibinfo {author}
		{\bibfnamefont {R.}~\bibnamefont {Langer}},\ }\bibfield  {title} {\bibinfo
		{title} {How to control defect formation in monolithic iii/v hetero-epitaxy
			on (100) si? a critical review on current approaches},\ }\href
	{https://doi.org/10.1088/1361-6641/aad655} {\bibfield  {journal} {\bibinfo
			{journal} {Semiconductor Science and Technology}\ }\textbf {\bibinfo {volume}
			{33}},\ \bibinfo {pages} {093002} (\bibinfo {year} {2018})}\BibitemShut
	{NoStop}%
	\bibitem [{\citenamefont {Benyoucef}\ \emph
		{et~al.}(2013{\natexlab{a}})\citenamefont {Benyoucef}, \citenamefont
		{Alzoubi}, \citenamefont {Reithmaier}, \citenamefont {Wu},\ and\
		\citenamefont {Trampert}}]{Benyoucef2013}%
	\BibitemOpen
	\bibfield  {author} {\bibinfo {author} {\bibfnamefont {M.}~\bibnamefont
			{Benyoucef}}, \bibinfo {author} {\bibfnamefont {T.}~\bibnamefont {Alzoubi}},
		\bibinfo {author} {\bibfnamefont {J.~P.}\ \bibnamefont {Reithmaier}},
		\bibinfo {author} {\bibfnamefont {M.}~\bibnamefont {Wu}},\ and\ \bibinfo
		{author} {\bibfnamefont {A.}~\bibnamefont {Trampert}},\ }\bibfield  {title}
	{\bibinfo {title} {Nanostructured hybrid material based on highly mismatched
			{III-V} nanocrystals fully embedded in silicon},\ }\href
	{https://doi.org/10.1002/pssa.201330395} {\bibfield  {journal} {\bibinfo
			{journal} {Phys. Stat. Solidi (a)}\ }\textbf {\bibinfo {volume} {211}},\
		\bibinfo {pages} {817–822} (\bibinfo {year}
		{2013}{\natexlab{a}})}\BibitemShut {NoStop}%
	\bibitem [{\citenamefont {Benyoucef}\ \emph
		{et~al.}(2013{\natexlab{b}})\citenamefont {Benyoucef}, \citenamefont
		{Usman},\ and\ \citenamefont {Reithmaier}}]{Benyoucef2013b}%
	\BibitemOpen
	\bibfield  {author} {\bibinfo {author} {\bibfnamefont {M.}~\bibnamefont
			{Benyoucef}}, \bibinfo {author} {\bibfnamefont {M.}~\bibnamefont {Usman}},\
		and\ \bibinfo {author} {\bibfnamefont {J.~P.}\ \bibnamefont {Reithmaier}},\
	}\bibfield  {title} {\bibinfo {title} {Bright light emissions with narrow
			spectral linewidths from single inas/gaas quantum dots directly grown on
			silicon substrates},\ }\href {https://doi.org/10.1063/1.4799149} {\bibfield
		{journal} {\bibinfo  {journal} {Appl. Phys. Lett.}\ }\textbf {\bibinfo
			{volume} {102}},\ \bibinfo {pages} {132101} (\bibinfo {year}
		{2013}{\natexlab{b}})}\BibitemShut {NoStop}%
	\bibitem [{\citenamefont {Wu}\ \emph {et~al.}(2015)\citenamefont {Wu},
		\citenamefont {Trampert}, \citenamefont {Al-Zoubi}, \citenamefont
		{Benyoucef},\ and\ \citenamefont {Reithmaier}}]{Wu2015}%
	\BibitemOpen
	\bibfield  {author} {\bibinfo {author} {\bibfnamefont {M.}~\bibnamefont
			{Wu}}, \bibinfo {author} {\bibfnamefont {A.}~\bibnamefont {Trampert}},
		\bibinfo {author} {\bibfnamefont {T.}~\bibnamefont {Al-Zoubi}}, \bibinfo
		{author} {\bibfnamefont {M.}~\bibnamefont {Benyoucef}},\ and\ \bibinfo
		{author} {\bibfnamefont {J.~P.}\ \bibnamefont {Reithmaier}},\ }\bibfield
	{title} {\bibinfo {title} {Interface structure and strain state of inas
			nano-clusters embedded in silicon},\ }\href
	{https://doi.org/10.1016/j.actamat.2015.02.042} {\bibfield  {journal}
		{\bibinfo  {journal} {Acta Materialia}\ }\textbf {\bibinfo {volume} {90}},\
		\bibinfo {pages} {133–139} (\bibinfo {year} {2015})}\BibitemShut {NoStop}%
	\bibitem [{\citenamefont {Keating}(1966)}]{Keating1966}%
	\BibitemOpen
	\bibfield  {author} {\bibinfo {author} {\bibfnamefont {P.~N.}\ \bibnamefont
			{Keating}},\ }\bibfield  {title} {\bibinfo {title} {Effect of {{Invariance
					Requirements}} on the {{Elastic Strain Energy}} of {{Crystals}} with
			{{Application}} to the {{Diamond Structure}}},\ }\href
	{https://doi.org/10.1103/PhysRev.145.637} {\bibfield  {journal} {\bibinfo
			{journal} {Phys. Rev.}\ }\textbf {\bibinfo {volume} {145}},\ \bibinfo {pages}
		{637} (\bibinfo {year} {1966})}\BibitemShut {NoStop}%
	\bibitem [{\citenamefont {Pryor}\ \emph {et~al.}(1998)\citenamefont {Pryor},
		\citenamefont {Kim}, \citenamefont {Wang}, \citenamefont {Williamson},\ and\
		\citenamefont {Zunger}}]{pryor98b}%
	\BibitemOpen
	\bibfield  {author} {\bibinfo {author} {\bibfnamefont {C.}~\bibnamefont
			{Pryor}}, \bibinfo {author} {\bibfnamefont {J.}~\bibnamefont {Kim}}, \bibinfo
		{author} {\bibfnamefont {L.~W.}\ \bibnamefont {Wang}}, \bibinfo {author}
		{\bibfnamefont {A.~J.}\ \bibnamefont {Williamson}},\ and\ \bibinfo {author}
		{\bibfnamefont {A.}~\bibnamefont {Zunger}},\ }\bibfield  {title} {\bibinfo
		{title} {{Comparison of two methods for describing the strain profiles in
				quantum dots}},\ }\href@noop {} {\bibfield  {journal} {\bibinfo  {journal}
			{J. Appl. Phys.}\ }\textbf {\bibinfo {volume} {83}},\ \bibinfo {pages} {2548}
		(\bibinfo {year} {1998})}\BibitemShut {NoStop}%
	\bibitem [{\citenamefont {Slater}\ and\ \citenamefont
		{Koster}(1954)}]{Slater1954}%
	\BibitemOpen
	\bibfield  {author} {\bibinfo {author} {\bibfnamefont {J.~C.}\ \bibnamefont
			{Slater}}\ and\ \bibinfo {author} {\bibfnamefont {G.~F.}\ \bibnamefont
			{Koster}},\ }\bibfield  {title} {\bibinfo {title} {{Simplified LCAO Method
				for the Periodic Potential Problem}},\ }\href
	{https://doi.org/10.1103/PhysRev.94.1498} {\bibfield  {journal} {\bibinfo
			{journal} {Phys. Rev.}\ }\textbf {\bibinfo {volume} {94}},\ \bibinfo {pages}
		{1498} (\bibinfo {year} {1954})}\BibitemShut {NoStop}%
	\bibitem [{\citenamefont {Jancu}\ \emph {et~al.}(1998)\citenamefont {Jancu},
		\citenamefont {Scholz}, \citenamefont {Beltram},\ and\ \citenamefont
		{Bassani}}]{Jancu1998}%
	\BibitemOpen
	\bibfield  {author} {\bibinfo {author} {\bibfnamefont {J.-M.}\ \bibnamefont
			{Jancu}}, \bibinfo {author} {\bibfnamefont {R.}~\bibnamefont {Scholz}},
		\bibinfo {author} {\bibfnamefont {F.}~\bibnamefont {Beltram}},\ and\ \bibinfo
		{author} {\bibfnamefont {F.}~\bibnamefont {Bassani}},\ }\bibfield  {title}
	{\bibinfo {title} {{Empirical tight-binding calculation for cubic
				semiconductors: General method and material parameters}},\ }\href
	{https://doi.org/10.1103/PhysRevB.57.6493} {\bibfield  {journal} {\bibinfo
			{journal} {Phys. Rev. B}\ }\textbf {\bibinfo {volume} {57}},\ \bibinfo
		{pages} {6493} (\bibinfo {year} {1998})}\BibitemShut {NoStop}%
	\bibitem [{\citenamefont {Zieli{\'{n}}ski}(2012{\natexlab{a}})}]{zielinski12}%
	\BibitemOpen
	\bibfield  {author} {\bibinfo {author} {\bibfnamefont {M.}~\bibnamefont
			{Zieli{\'{n}}ski}},\ }\bibfield  {title} {\bibinfo {title} {{Including strain
				in atomistic tight-binding Hamiltonians: An application to self-assembled
				InAs/GaAs and InAs/InP quantum dots}},\ }\href
	{https://doi.org/10.1103/PhysRevB.86.115424} {\bibfield  {journal} {\bibinfo
			{journal} {Phys. Rev. B}\ }\textbf {\bibinfo {volume} {86}},\ \bibinfo
		{pages} {115424} (\bibinfo {year} {2012}{\natexlab{a}})}\BibitemShut
	{NoStop}%
	\bibitem [{\citenamefont {Chadi}(1977)}]{Chadi1977}%
	\BibitemOpen
	\bibfield  {author} {\bibinfo {author} {\bibfnamefont {D.~J.}\ \bibnamefont
			{Chadi}},\ }\bibfield  {title} {\bibinfo {title} {{Spin-orbit splitting in
				crystalline and compositionally disordered semiconductors}},\ }\href
	{https://doi.org/10.1103/PhysRevB.16.790} {\bibfield  {journal} {\bibinfo
			{journal} {Phys. Rev. B}\ }\textbf {\bibinfo {volume} {16}},\ \bibinfo
		{pages} {790} (\bibinfo {year} {1977})}\BibitemShut {NoStop}%
	\bibitem [{\citenamefont {Ren}\ \emph {et~al.}(1982)\citenamefont {Ren},
		\citenamefont {Dow},\ and\ \citenamefont {Wolford}}]{Ren1982}%
	\BibitemOpen
	\bibfield  {author} {\bibinfo {author} {\bibfnamefont {S.~Y.}\ \bibnamefont
			{Ren}}, \bibinfo {author} {\bibfnamefont {J.~D.}\ \bibnamefont {Dow}},\ and\
		\bibinfo {author} {\bibfnamefont {D.~J.}\ \bibnamefont {Wolford}},\
	}\bibfield  {title} {\bibinfo {title} {Pressure dependence of deep levels in
			gaas},\ }\href {https://doi.org/10.1103/PhysRevB.25.7661} {\bibfield
		{journal} {\bibinfo  {journal} {Phys. Rev. B}\ }\textbf {\bibinfo {volume}
			{25}},\ \bibinfo {pages} {7661} (\bibinfo {year} {1982})}\BibitemShut
	{NoStop}%
	\bibitem [{\citenamefont {Jancu}\ and\ \citenamefont
		{Voisin}(2007)}]{Jancu2007}%
	\BibitemOpen
	\bibfield  {author} {\bibinfo {author} {\bibfnamefont {J.-M.}\ \bibnamefont
			{Jancu}}\ and\ \bibinfo {author} {\bibfnamefont {P.}~\bibnamefont {Voisin}},\
	}\bibfield  {title} {\bibinfo {title} {{Tetragonal and trigonal deformations
				in zinc-blende semiconductors: A tight-binding point of view}},\ }\href
	{https://doi.org/10.1103/PhysRevB.76.115202} {\bibfield  {journal} {\bibinfo
			{journal} {Phys. Rev. B}\ }\textbf {\bibinfo {volume} {76}},\ \bibinfo
		{pages} {115202} (\bibinfo {year} {2007})}\BibitemShut {NoStop}%
	\bibitem [{\citenamefont {Gawarecki}\ and\ \citenamefont
		{Zieli{\'{n}}ski}(2019)}]{Gawarecki2019b}%
	\BibitemOpen
	\bibfield  {author} {\bibinfo {author} {\bibfnamefont {K.}~\bibnamefont
			{Gawarecki}}\ and\ \bibinfo {author} {\bibfnamefont {M.}~\bibnamefont
			{Zieli{\'{n}}ski}},\ }\bibfield  {title} {\bibinfo {title} {{Importance of
				second-order deformation potentials in modeling of InAs/GaAs
				nanostructures}},\ }\href {https://doi.org/10.1103/PhysRevB.100.155409}
	{\bibfield  {journal} {\bibinfo  {journal} {Phys. Rev. B}\ }\textbf {\bibinfo
			{volume} {100}},\ \bibinfo {pages} {155409} (\bibinfo {year}
		{2019})}\BibitemShut {NoStop}%
	\bibitem [{\citenamefont {Feynman}(1939)}]{Feynman1939}%
	\BibitemOpen
	\bibfield  {author} {\bibinfo {author} {\bibfnamefont {R.~P.}\ \bibnamefont
			{Feynman}},\ }\bibfield  {title} {\bibinfo {title} {{Forces in Molecules}},\
	}\href {https://doi.org/10.1103/PhysRev.56.340} {\bibfield  {journal}
		{\bibinfo  {journal} {Phys. Rev.}\ }\textbf {\bibinfo {volume} {56}},\
		\bibinfo {pages} {340} (\bibinfo {year} {1939})}\BibitemShut {NoStop}%
	\bibitem [{\citenamefont {{Lew Yan Voon}}\ and\ \citenamefont
		{Ram-Mohan}(1993)}]{LewYanVoon1993}%
	\BibitemOpen
	\bibfield  {author} {\bibinfo {author} {\bibfnamefont {L.~C.}\ \bibnamefont
			{{Lew Yan Voon}}}\ and\ \bibinfo {author} {\bibfnamefont {L.~R.}\
			\bibnamefont {Ram-Mohan}},\ }\bibfield  {title} {\bibinfo {title}
		{{Tight-binding representation of the optical matrix elements: Theory and
				applications}},\ }\href {https://doi.org/10.1103/PhysRevB.47.15500}
	{\bibfield  {journal} {\bibinfo  {journal} {Phys. Rev. B}\ }\textbf {\bibinfo
			{volume} {47}},\ \bibinfo {pages} {15500} (\bibinfo {year}
		{1993})}\BibitemShut {NoStop}%
	\bibitem [{\citenamefont {Gawarecki}\ and\ \citenamefont
		{Zieli{\'{n}}ski}(2020)}]{Gawarecki2020}%
	\BibitemOpen
	\bibfield  {author} {\bibinfo {author} {\bibfnamefont {K.}~\bibnamefont
			{Gawarecki}}\ and\ \bibinfo {author} {\bibfnamefont {M.}~\bibnamefont
			{Zieli{\'{n}}ski}},\ }\bibfield  {title} {\bibinfo {title} {{Electron
				g-factor in nanostructures: continuum media and atomistic approach}},\ }\href
	{https://doi.org/10.1038/s41598-020-79133-0} {\bibfield  {journal} {\bibinfo
			{journal} {Sci. Rep.}\ }\textbf {\bibinfo {volume} {10}},\ \bibinfo {pages}
		{22001} (\bibinfo {year} {2020})}\BibitemShut {NoStop}%
	\bibitem [{\citenamefont {Bryant}(1987)}]{Bryant1987}%
	\BibitemOpen
	\bibfield  {author} {\bibinfo {author} {\bibfnamefont {G.~W.}\ \bibnamefont
			{Bryant}},\ }\bibfield  {title} {\bibinfo {title} {Electronic structure of
			ultrasmall quantum-well boxes},\ }\href
	{https://doi.org/10.1103/PhysRevLett.59.1140} {\bibfield  {journal} {\bibinfo
			{journal} {Phys. Rev. Lett.}\ }\textbf {\bibinfo {volume} {59}},\ \bibinfo
		{pages} {1140} (\bibinfo {year} {1987})}\BibitemShut {NoStop}%
	\bibitem [{\citenamefont {Gawarecki}\ \emph {et~al.}(2024)\citenamefont
		{Gawarecki}, \citenamefont {Ziembicki}, \citenamefont {Scharoch},\ and\
		\citenamefont {Kudrawiec}}]{Gawarecki2024}%
	\BibitemOpen
	\bibfield  {author} {\bibinfo {author} {\bibfnamefont {K.}~\bibnamefont
			{Gawarecki}}, \bibinfo {author} {\bibfnamefont {J.}~\bibnamefont
			{Ziembicki}}, \bibinfo {author} {\bibfnamefont {P.}~\bibnamefont
			{Scharoch}},\ and\ \bibinfo {author} {\bibfnamefont {R.}~\bibnamefont
			{Kudrawiec}},\ }\bibfield  {title} {\bibinfo {title} {Electronic and spectral
			properties of {Ge$_{1-x}$Sn$_x$} quantum dots},\ }\href
	{https://doi.org/10.1063/5.0198146} {\bibfield  {journal} {\bibinfo
			{journal} {J. Appl. Phys.}\ }\textbf {\bibinfo {volume} {135}},\ \bibinfo
		{pages} {214303} (\bibinfo {year} {2024})}\BibitemShut {NoStop}%
	\bibitem [{\citenamefont {Choi}\ \emph {et~al.}(2013)\citenamefont {Choi},
		\citenamefont {Lyons}, \citenamefont {Janotti},\ and\ \citenamefont {Van~de
			Walle}}]{Choi2013}%
	\BibitemOpen
	\bibfield  {author} {\bibinfo {author} {\bibfnamefont {M.}~\bibnamefont
			{Choi}}, \bibinfo {author} {\bibfnamefont {J.~L.}\ \bibnamefont {Lyons}},
		\bibinfo {author} {\bibfnamefont {A.}~\bibnamefont {Janotti}},\ and\ \bibinfo
		{author} {\bibfnamefont {C.~G.}\ \bibnamefont {Van~de Walle}},\ }\bibfield
	{title} {\bibinfo {title} {Impact of carbon and nitrogen impurities in
			high-kappa dielectrics on metal-oxide-semiconductor devices},\ }\href
	{https://doi.org/10.1063/1.4801497} {\bibfield  {journal} {\bibinfo
			{journal} {Appl. Phys. Lett.}\ }\textbf {\bibinfo {volume} {102}},\ \bibinfo
		{pages} {142902} (\bibinfo {year} {2013})}\BibitemShut {NoStop}%
	\bibitem [{\citenamefont {Krawicz}\ \emph {et~al.}(2014)\citenamefont
		{Krawicz}, \citenamefont {Cedeno},\ and\ \citenamefont
		{Moore}}]{Krawicz2014}%
	\BibitemOpen
	\bibfield  {author} {\bibinfo {author} {\bibfnamefont {A.}~\bibnamefont
			{Krawicz}}, \bibinfo {author} {\bibfnamefont {D.}~\bibnamefont {Cedeno}},\
		and\ \bibinfo {author} {\bibfnamefont {G.~F.}\ \bibnamefont {Moore}},\
	}\bibfield  {title} {\bibinfo {title} {Energetics and efficiency analysis of
			a cobaloxime-modified semiconductor under simulated air mass 1.5
			illumination},\ }\href {https://doi.org/10.1039/c4cp00495g} {\bibfield
		{journal} {\bibinfo  {journal} {Phys. Chem. Chem. Phys.}\ }\textbf {\bibinfo
			{volume} {16}},\ \bibinfo {pages} {15818–15824} (\bibinfo {year}
		{2014})}\BibitemShut {NoStop}%
	\bibitem [{\citenamefont {Kang}\ \emph {et~al.}(2015)\citenamefont {Kang},
		\citenamefont {Kim}, \citenamefont {Kubota}, \citenamefont {Cardiel},
		\citenamefont {Cha},\ and\ \citenamefont {Choi}}]{Kang2015}%
	\BibitemOpen
	\bibfield  {author} {\bibinfo {author} {\bibfnamefont {D.}~\bibnamefont
			{Kang}}, \bibinfo {author} {\bibfnamefont {T.~W.}\ \bibnamefont {Kim}},
		\bibinfo {author} {\bibfnamefont {S.~R.}\ \bibnamefont {Kubota}}, \bibinfo
		{author} {\bibfnamefont {A.~C.}\ \bibnamefont {Cardiel}}, \bibinfo {author}
		{\bibfnamefont {H.~G.}\ \bibnamefont {Cha}},\ and\ \bibinfo {author}
		{\bibfnamefont {K.-S.}\ \bibnamefont {Choi}},\ }\bibfield  {title} {\bibinfo
		{title} {Electrochemical synthesis of photoelectrodes and catalysts for use
			in solar water splitting},\ }\href
	{https://doi.org/10.1021/acs.chemrev.5b00498} {\bibfield  {journal} {\bibinfo
			{journal} {Chemical Reviews}\ }\textbf {\bibinfo {volume} {115}},\ \bibinfo
		{pages} {12839–12887} (\bibinfo {year} {2015})}\BibitemShut {NoStop}%
	\bibitem [{\citenamefont {Vurgaftman}\ \emph {et~al.}(2001)\citenamefont
		{Vurgaftman}, \citenamefont {Meyer},\ and\ \citenamefont
		{Ram-Mohan}}]{vurgaftman01}%
	\BibitemOpen
	\bibfield  {author} {\bibinfo {author} {\bibfnamefont {I.}~\bibnamefont
			{Vurgaftman}}, \bibinfo {author} {\bibfnamefont {J.~R.}\ \bibnamefont
			{Meyer}},\ and\ \bibinfo {author} {\bibfnamefont {L.~R.}\ \bibnamefont
			{Ram-Mohan}},\ }\bibfield  {title} {\bibinfo {title} {{Band parameters for
				III-V compound semiconductors and their alloys}},\ }\href
	{https://doi.org/10.1063/1.1368156} {\bibfield  {journal} {\bibinfo
			{journal} {J. Appl. Phys.}\ }\textbf {\bibinfo {volume} {89}},\ \bibinfo
		{pages} {5815} (\bibinfo {year} {2001})}\BibitemShut {NoStop}%
	\bibitem [{\citenamefont
		{Zieli{\'{n}}ski}(2012{\natexlab{b}})}]{Zielinski2012}%
	\BibitemOpen
	\bibfield  {author} {\bibinfo {author} {\bibfnamefont {M.}~\bibnamefont
			{Zieli{\'{n}}ski}},\ }\bibfield  {title} {\bibinfo {title} {{Influence of
				substrate orientation on exciton fine structure splitting of InAs/InP
				nanowire quantum dots}},\ }\href {https://doi.org/10.1186/1556-276X-7-265}
	{\bibfield  {journal} {\bibinfo  {journal} {Nanoscale Res. Lett.}\ }\textbf
		{\bibinfo {volume} {7}},\ \bibinfo {pages} {265} (\bibinfo {year}
		{2012}{\natexlab{b}})}\BibitemShut {NoStop}%
	\bibitem [{\citenamefont {Balay}\ \emph {et~al.}(2024)\citenamefont {Balay},
		\citenamefont {Abhyankar}, \citenamefont {Adams}, \citenamefont {Benson},
		\citenamefont {Brown}, \citenamefont {Brune}, \citenamefont {Buschelman},
		\citenamefont {Constantinescu}, \citenamefont {Dalcin}, \citenamefont
		{Dener}, \citenamefont {Eijkhout}, \citenamefont {Faibussowitsch},
		\citenamefont {Gropp}, \citenamefont {Hapla}, \citenamefont {Isaac},
		\citenamefont {Jolivet}, \citenamefont {Karpeev}, \citenamefont {Kaushik},
		\citenamefont {Knepley}, \citenamefont {Kong}, \citenamefont {Kruger},
		\citenamefont {May}, \citenamefont {McInnes}, \citenamefont {Mills},
		\citenamefont {Mitchell}, \citenamefont {Munson}, \citenamefont {Roman},
		\citenamefont {Rupp}, \citenamefont {Sanan}, \citenamefont {Sarich},
		\citenamefont {Smith}, \citenamefont {Zampini}, \citenamefont {Zhang},
		\citenamefont {Zhang},\ and\ \citenamefont {Zhang}}]{petsc-web-page}%
	\BibitemOpen
	\bibfield  {author} {\bibinfo {author} {\bibfnamefont {S.}~\bibnamefont
			{Balay}}, \bibinfo {author} {\bibfnamefont {S.}~\bibnamefont {Abhyankar}},
		\bibinfo {author} {\bibfnamefont {M.~F.}\ \bibnamefont {Adams}}, \bibinfo
		{author} {\bibfnamefont {S.}~\bibnamefont {Benson}}, \bibinfo {author}
		{\bibfnamefont {J.}~\bibnamefont {Brown}}, \bibinfo {author} {\bibfnamefont
			{P.}~\bibnamefont {Brune}}, \bibinfo {author} {\bibfnamefont
			{K.}~\bibnamefont {Buschelman}}, \bibinfo {author} {\bibfnamefont {E.~M.}\
			\bibnamefont {Constantinescu}}, \bibinfo {author} {\bibfnamefont
			{L.}~\bibnamefont {Dalcin}}, \bibinfo {author} {\bibfnamefont
			{A.}~\bibnamefont {Dener}}, \bibinfo {author} {\bibfnamefont
			{V.}~\bibnamefont {Eijkhout}}, \bibinfo {author} {\bibfnamefont
			{J.}~\bibnamefont {Faibussowitsch}}, \bibinfo {author} {\bibfnamefont
			{W.~D.}\ \bibnamefont {Gropp}}, \bibinfo {author} {\bibfnamefont
			{V.}~\bibnamefont {Hapla}}, \bibinfo {author} {\bibfnamefont
			{T.}~\bibnamefont {Isaac}}, \bibinfo {author} {\bibfnamefont
			{P.}~\bibnamefont {Jolivet}}, \bibinfo {author} {\bibfnamefont
			{D.}~\bibnamefont {Karpeev}}, \bibinfo {author} {\bibfnamefont
			{D.}~\bibnamefont {Kaushik}}, \bibinfo {author} {\bibfnamefont {M.~G.}\
			\bibnamefont {Knepley}}, \bibinfo {author} {\bibfnamefont {F.}~\bibnamefont
			{Kong}}, \bibinfo {author} {\bibfnamefont {S.}~\bibnamefont {Kruger}},
		\bibinfo {author} {\bibfnamefont {D.~A.}\ \bibnamefont {May}}, \bibinfo
		{author} {\bibfnamefont {L.~C.}\ \bibnamefont {McInnes}}, \bibinfo {author}
		{\bibfnamefont {R.~T.}\ \bibnamefont {Mills}}, \bibinfo {author}
		{\bibfnamefont {L.}~\bibnamefont {Mitchell}}, \bibinfo {author}
		{\bibfnamefont {T.}~\bibnamefont {Munson}}, \bibinfo {author} {\bibfnamefont
			{J.~E.}\ \bibnamefont {Roman}}, \bibinfo {author} {\bibfnamefont
			{K.}~\bibnamefont {Rupp}}, \bibinfo {author} {\bibfnamefont {P.}~\bibnamefont
			{Sanan}}, \bibinfo {author} {\bibfnamefont {J.}~\bibnamefont {Sarich}},
		\bibinfo {author} {\bibfnamefont {B.~F.}\ \bibnamefont {Smith}}, \bibinfo
		{author} {\bibfnamefont {S.}~\bibnamefont {Zampini}}, \bibinfo {author}
		{\bibfnamefont {H.}~\bibnamefont {Zhang}}, \bibinfo {author} {\bibfnamefont
			{H.}~\bibnamefont {Zhang}},\ and\ \bibinfo {author} {\bibfnamefont
			{J.}~\bibnamefont {Zhang}},\ }\href@noop {} {\bibinfo {title} {{{PETSc Web}}
			page}} (\bibinfo {year} {2024})\BibitemShut {NoStop}%
	\bibitem [{\citenamefont {Lee}\ \emph {et~al.}(2004)\citenamefont {Lee},
		\citenamefont {Oyafuso}, \citenamefont {von Allmen},\ and\ \citenamefont
		{Klimeck}}]{Lee2004}%
	\BibitemOpen
	\bibfield  {author} {\bibinfo {author} {\bibfnamefont {S.}~\bibnamefont
			{Lee}}, \bibinfo {author} {\bibfnamefont {F.}~\bibnamefont {Oyafuso}},
		\bibinfo {author} {\bibfnamefont {P.}~\bibnamefont {von Allmen}},\ and\
		\bibinfo {author} {\bibfnamefont {G.}~\bibnamefont {Klimeck}},\ }\bibfield
	{title} {\bibinfo {title} {{Boundary conditions for the electronic structure
				of finite-extent embedded semiconductor nanostructures}},\ }\href
	{https://doi.org/10.1103/PhysRevB.69.045316} {\bibfield  {journal} {\bibinfo
			{journal} {Phys. Rev. B}\ }\textbf {\bibinfo {volume} {69}},\ \bibinfo
		{pages} {45316} (\bibinfo {year} {2004})}\BibitemShut {NoStop}%
	\bibitem [{\citenamefont {Eissfeller}(2012)}]{eissfeller12}%
	\BibitemOpen
	\bibfield  {author} {\bibinfo {author} {\bibfnamefont {T.}~\bibnamefont
			{Eissfeller}},\ }\emph {\bibinfo {title} {{Theory of the Electronic Structure
				of Quantum Dots in External Fields}}},\ \href@noop {} {Ph.D. thesis},\
	\bibinfo  {school} {Technical University of Munich} (\bibinfo {year}
	{2012})\BibitemShut {NoStop}%
	\bibitem [{\citenamefont {Thr{\"a}nhardt}\ \emph {et~al.}(2002)\citenamefont
		{Thr{\"a}nhardt}, \citenamefont {Ell}, \citenamefont {Khitrova},\ and\
		\citenamefont {Gibbs}}]{Thranhardt2002}%
	\BibitemOpen
	\bibfield  {author} {\bibinfo {author} {\bibfnamefont {A.}~\bibnamefont
			{Thr{\"a}nhardt}}, \bibinfo {author} {\bibfnamefont {C.}~\bibnamefont {Ell}},
		\bibinfo {author} {\bibfnamefont {G.}~\bibnamefont {Khitrova}},\ and\
		\bibinfo {author} {\bibfnamefont {H.~M.}\ \bibnamefont {Gibbs}},\ }\bibfield
	{title} {\bibinfo {title} {Relation between dipole moment and radiative
			lifetime in interface fluctuation quantum dots},\ }\href
	{https://doi.org/10.1103/PhysRevB.65.035327} {\bibfield  {journal} {\bibinfo
			{journal} {Physical Review B}\ }\textbf {\bibinfo {volume} {65}},\ \bibinfo
		{pages} {035327} (\bibinfo {year} {2002})}\BibitemShut {NoStop}%
	\bibitem [{\citenamefont {Gawe{\l}czyk}\ \emph {et~al.}(2017)\citenamefont
		{Gawe{\l}czyk}, \citenamefont {Syperek}, \citenamefont {Mary{\'n}ski},
		\citenamefont {Mrowi{\'n}ski}, \citenamefont {Dusanowski}, \citenamefont
		{Gawarecki}, \citenamefont {Misiewicz}, \citenamefont {Somers}, \citenamefont
		{Reithmaier}, \citenamefont {H{\"o}fling},\ and\ \citenamefont {S{\k
				e}k}}]{Gawelczyk2017}%
	\BibitemOpen
	\bibfield  {author} {\bibinfo {author} {\bibfnamefont {M.}~\bibnamefont
			{Gawe{\l}czyk}}, \bibinfo {author} {\bibfnamefont {M.}~\bibnamefont
			{Syperek}}, \bibinfo {author} {\bibfnamefont {A.}~\bibnamefont
			{Mary{\'n}ski}}, \bibinfo {author} {\bibfnamefont {P.}~\bibnamefont
			{Mrowi{\'n}ski}}, \bibinfo {author} {\bibfnamefont {{\L}.}~\bibnamefont
			{Dusanowski}}, \bibinfo {author} {\bibfnamefont {K.}~\bibnamefont
			{Gawarecki}}, \bibinfo {author} {\bibfnamefont {J.}~\bibnamefont
			{Misiewicz}}, \bibinfo {author} {\bibfnamefont {A.}~\bibnamefont {Somers}},
		\bibinfo {author} {\bibfnamefont {J.~P.}\ \bibnamefont {Reithmaier}},
		\bibinfo {author} {\bibfnamefont {S.}~\bibnamefont {H{\"o}fling}},\ and\
		\bibinfo {author} {\bibfnamefont {G.}~\bibnamefont {S{\k e}k}},\ }\bibfield
	{title} {\bibinfo {title} {Exciton lifetime and emission polarization
			dispersion in strongly in-plane asymmetric nanostructures},\ }\href
	{https://doi.org/10.1103/PhysRevB.96.245425} {\bibfield  {journal} {\bibinfo
			{journal} {Phys. Rev. B}\ }\textbf {\bibinfo {volume} {96}},\ \bibinfo
		{pages} {245425} (\bibinfo {year} {2017})}\BibitemShut {NoStop}%
\end{thebibliography}
\end{document}